\documentclass[12pt]{article}

\usepackage[dvips]{graphicx}
\usepackage{amsmath,amssymb,amsthm}
\usepackage{setspace}
\usepackage{hyperref}
\usepackage{mwe}
\usepackage{subfig}
\usepackage{natbib}
\usepackage{color}
\usepackage{array}
\usepackage{float}
\usepackage{endfloat}

\begin{document}


\title{\bf Regional Synchronization during Economic Contraction: The Case of the U.S. and Japan}

\author{Makoto Muto~
\thanks{Kanda-Misakicho 1--3--2, Chiyoda-ku, Tokyo 101--8360, Japan.
\textit{E-mail}: muto.makoto@nihon-u.ac.jp.}\\
{\scriptsize College of Economics, Nihon University}\\
\and Tamotsu Onozaki~
\thanks{Corresponding author. Osaki 4--2--16, Shinagawa-ku, Tokyo 141--8602, Japan. \textit{E-mail}: onozaki@ris.ac.jp.}\\
{\scriptsize Faculty of Economics, Rissho University}\\
\and Yoshitaka Saiki~
\thanks{Naka 2--1, Kunitachi-shi, Tokyo 186--8601, Japan. \textit{E-mail}: yoshi.saiki@r.hit-u.ac.jp}\\
{\scriptsize Graduate School of Business Administration, Hitotsubashi University}
}

\date{}

\maketitle


\newpage
\begin{abstract}

\noindent
Two decades of studies have found significant regional differences in the timing of transitions in national business cycles and their durations.
Earlier studies partly detect regional synchronization during business cycle expansions and contractions in Europe, the United States, and Japan.
We examine this possibility applying a sophisticated method for identifying the time-varying degree of synchronization to regional business cycle data in the U.S. and Japan. 
The method is prominent in nonlinear sciences but has been infrequently applied in business cycle studies.
We find that synchronization in regional business cycles increased during contractions and decreased during expansions throughout the period under study.
Such asymmetry between the contraction and expansion phases of a business cycle will contribute our better understanding of the phenomenon of business cycles.

\vspace{4ex}

\vspace{2ex}

\end{abstract}

\newpage
\section{Introduction}
\label{intro}

National economies consist of interlinked regional economies that react differently to changing macroeconomic forces, government policies, imported materials prices, and technological innovation.
Thus, national business cycles (BCs) are an admixture of regional cycles fluctuating diversely.
Earlier studies of regional BCs surveyed by \cite{domazlicky_1980} examined how and why cycles differ.
By contrast, the advent of the Economic and Monetary Union (EMU) in Europe has renewed research interest in similarities and synchronization among EU states' BCs because synchronization facilitates intra-EMU's fiscal and monetary policies.
However, empirical studies often reach divergent conclusions \citep{massmann_2004, grayer_2007, dehaan_2008, montoya_2008}, likely because they use different raw data or methods for estimating cycles and gauging synchronization.
For instance, \cite{artis_1999} found that synchronization intensified during the European Exchange Rate Mechanism period (1973--1995), but \cite{massmann_2004} determined periods of synchronization and desynchronization using identical but updated data.
Meanwhile, after conducting a comprehensive study using six estimation methods and three measures of synchronization, \cite{kappler_2013} found little support for BC synchronization, and degrees of synchronization fluctuated over time.

Euro-area studies of synchronization aroused interest in regional cycles within a country, such as the United States (U.S.) and Japan.
For example, \cite{clark_2001} specified that BCs of nine U.S. Census regions are significantly more synchronized than those of EU countries.
Moreover, \cite{artis_2011} found that the degree of regional BC synchronization within Japan is strikingly higher than that in the U.S. and the Euro area.
These findings imply that national borders may dampen synchronization between regional BCs.

Numerous studies regarding the synchronization of regional BCs, whether they cover inter- or \textit{intra}-national BCs, found significant regional differences in the timing of BCs' transitions and duration.
Among them, several studies announced noteworthy results.
For instance, \cite{grayer_2007} determined a recurring pattern of declining synchronization during expansions in Europe.
\cite{hamilton_2012} and \cite{chung_2016} noted that co-movement across states characterizes BC contractions in the U.S.
Meanwhile, \cite{wall_2007} concluded that contractions tend to be experienced across most Japanese prefectures.

These results garnered via different datasets and methods, when being put all together, suggest that the degree of synchronization between regional BCs intensifies during contractions and diminishes during expansions.
We examine this possibility applying a sophisticated method for identifying the time-varying degree of synchronization to regional BC data in the U.S. and Japan. 
As noted above, national borders may dampen synchronization between regional BCs, and therefore this study concentrates on analyzing the synchronization of regional BCs within a single country as a first step.
The method is prominent in nonlinear sciences (e.g., \citealp{pikovsky_2001}:ch.~6) but has been infrequently applied in BC studies.\footnote{
Another way to identify synchronization in time series is through the wavelet transform. 
Examples of previous studies using the cross-spectrum of wavelet coefficients to gauge synchronization of BCs include \cite{Aguiar_2011} for EU countries and \cite{Aguiar_2017} for U.S. states.
}
Before describing the method, we note how to extract BCs from raw data.
Following recent studies, we acknowledge that BCs are relative to a trend and focus on its deviation.

For monthly observations of the Composite Index (CI) of coincident indicators in the U.S. and the Index of Industrial Production (IIP) in Japan, we employ a band-pass filter to extract time series indicating regional BCs in both countries.
The Hodrick--Prescott filter (\citealp{hodrick_1997}), a high-pass filter often used in economics, removes only trends with low frequencies.
The Baxter--King (BK) (\citealp{baxter_1999}) and Christiano--Fitzgerald (CF) (\citealp{christiano_2003}) band-pass filters are also frequent in the literature, but we employ the Fourier band-pass filter that is mathematically and computationally simpler.\footnote
{
\cite{ikeda_2013} also use this filter.
}

This study's method comprises the following three procedures.
First, we convert time series fluctuations into two-dimensional oscillations using the Hilbert transform.\footnote{
The Hilbert transform is discussed in detail in Section~\ref{Hilbert}.
It has been used occasionally in the economics literature (see \citealp{ikeda_2013}).
}
This enables us to identify ``phases'' of circular oscillations, defined as a position of a cyclically oscillating variable within one period. 
Converted oscillations include more information than the original one-dimensional time series and better assess synchronization.
Second, we take the ``phase difference'' between two cycles to indicate their synchronization.
Third, we use the phase difference to calculate a synchronization index that measures the constancy of the phase difference.
If the phase difference of two cycles is nearly constant over time, this index indicates a value near 1, and we designate this situation as (phase) synchronization.
In this sense, synchronization does not depend on the level of the phase difference but on its constancy.
Our use of this method supports the hypothesis that synchronization between regional BCs intensifies during contractions and diminishes during expansions both in the U.S. and Japan.

An overview of other synchronization measures distinguishes our method from others.
The most popular measure of synchronization, namely, Pearson correlation coefficient, provides in one number the degree of similarity between series over a sampled period. 
It measures static relations between the series, whereas synchronization is a dynamic phenomenon with a varying degree over time.
Meanwhile, the moving window correlations and new time-varying indexes overcome that deficiency.
However, as the \cite{european_2006} mentioned, correlation with a moving window is sensitive to the window's length.
\cite{mink_2012} proposed a multivariate, time-varying measure of synchronization based on an output gap.
It gauges the percentage of regions over time whose output gap has the same sign as that of the reference region.
However, this synchronization measure is nondifferentiable because of its absolute values, and graphs of calculated series exhibit numerous nonessential spikes.

By contrast, our method focuses on phase differences between two time series.
Calculated using phase differences, the synchronization index of \cite{rosenblum_2001} captures the time-dependent degree of synchronization even if phase differences between two time series are large.
The correlation coefficient fails to measure the degree of synchronization because its absolute value can be small in such a case. 

The study proceeds as follows. 
Section \ref{data} and \ref{measuring} describe this study's data and methods, respectively.
Section \ref{analysis} presents empirical results.
Section \ref{conclusion} concludes the paper.

\section{Data}
\label{data}

We employ two datasets frequently used to investigate the regional BCs. 
One is monthly, seasonally adjusted CI data (2007 average $=100$) in the U.S.,  spanning from April 1979 to April 2021 (505 months) for 50 states compiled by the Federal Reserve Bank of Philadelphia. 
The other is monthly raw (i.e., not seasonally adjusted) IIP data (2010 average $=100$) in Japan, spanning from January 1978 to August 2018 (488 months) for 47 prefectures compiled by the Ministry of Economy, Trade, and Industry.\footnote
{
The CI data for 50 states in the U.S. are from the website of the Federal Reserve Bank of Philadelphia. The IIP data for Japan's 47 prefectures are from NIKKEI NEEDS.}
It does not matter whether the data used in the analysis are seasonally adjusted or not because our settings of the band-pass filter can remove the high-frequency component corresponding to the seasonal variation.

\begin{figure}[p]
    \begin{center} 
        \subfloat{({\bf u1}) }{\includegraphics[clip, width=0.42\columnwidth,height=0.28\columnwidth]{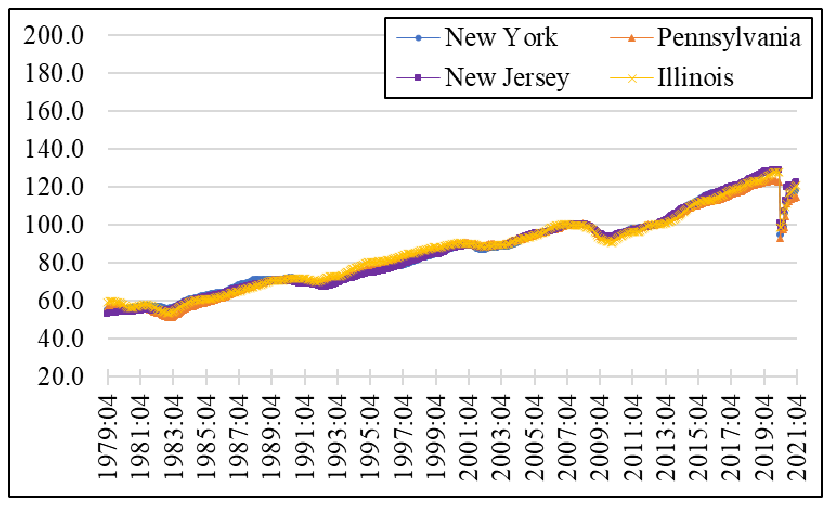}}
        \subfloat{({\bf j1}) }{\includegraphics[clip, width=0.42\columnwidth,height=0.28\columnwidth]{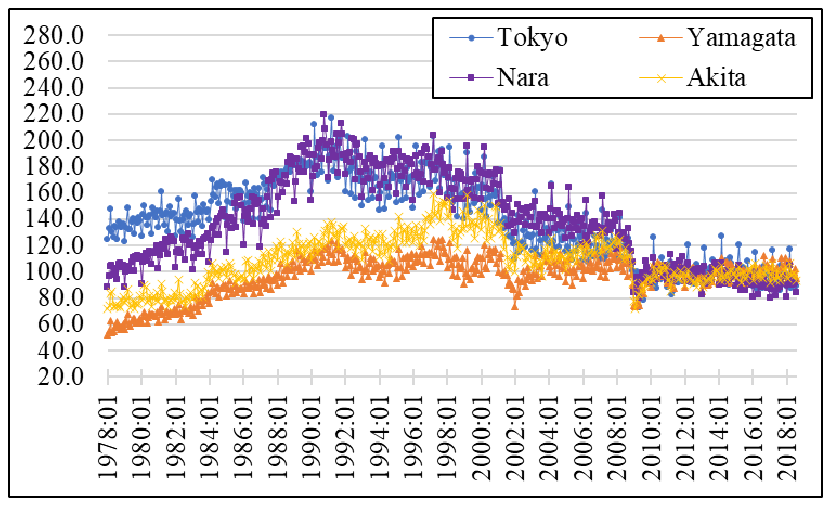}}\\
        \vspace{2mm}
        \subfloat{({\bf u2}) }{\includegraphics[clip, width=0.42\columnwidth,height=0.28\columnwidth]{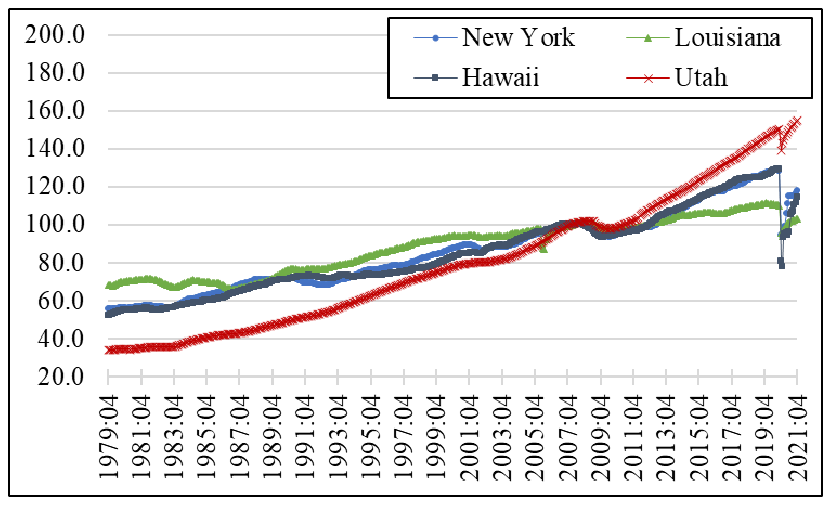}}
        \subfloat{({\bf j2}) }{\includegraphics[clip, width=0.42\columnwidth,height=0.28\columnwidth]{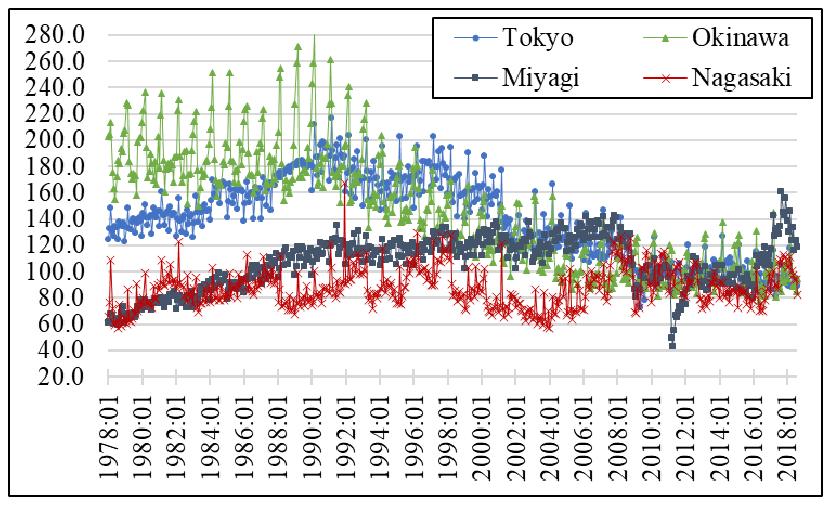}}\\
    \end{center}
    \caption{Time Series Comparison of the Composite Index and the Index of Industrial Production Data for Sampled Regions.\newline}
    \begin{spacing}{1.1}
        Note: Comparison of the composite index data between New York and other three states whose data exhibit the greatest synchronization with New York (i.e., Pennsylvania, New Jersey, and Illinois) (u1), and the least synchronization with New York (i.e., Louisiana, Hawaii, and Utah) (u2).
        Comparison of the index of industrial production data between Tokyo and other three prefectures whose data exhibit the greatest synchronization with Tokyo (i.e., Yamagata, Nara, and Akita) (j1), and the least synchronization with Tokyo (i.e., Okinawa, Miyagi, and Nagasaki) (j2).
    \end{spacing}

    \label{fig:time-series-iip}
\end{figure}

Figure~\ref{fig:time-series-iip} graphs the time series of CI and IIP data for sampled regions.
In particular, Figure~\ref{fig:time-series-iip}~(u1) compares the time series of CI in New York with those in Pennsylvania, New Jersey, and Illinois, the data for which exhibit the greatest synchronization with New York from the viewpoint of our analysis.
By contrast, Figure~\ref{fig:time-series-iip}~(u2) compares time series in New York with those in Louisiana, Hawaii, and Utah, the data for which exhibit the least synchronization with New York.
Meanwhile, Figure~\ref{fig:time-series-iip}~(j1) compares the time series of IIP in Tokyo with those in Yamagata, Nara, and Akita Prefectures, which exhibit the greatest synchronization with Tokyo.
Figure~\ref{fig:time-series-iip}~(j2) compares the time series in Tokyo with those in Okinawa, Miyagi, and Nagasaki Prefectures, showing the least synchronization with Tokyo.

\begin{figure}[p]
    \begin{center}
        \includegraphics{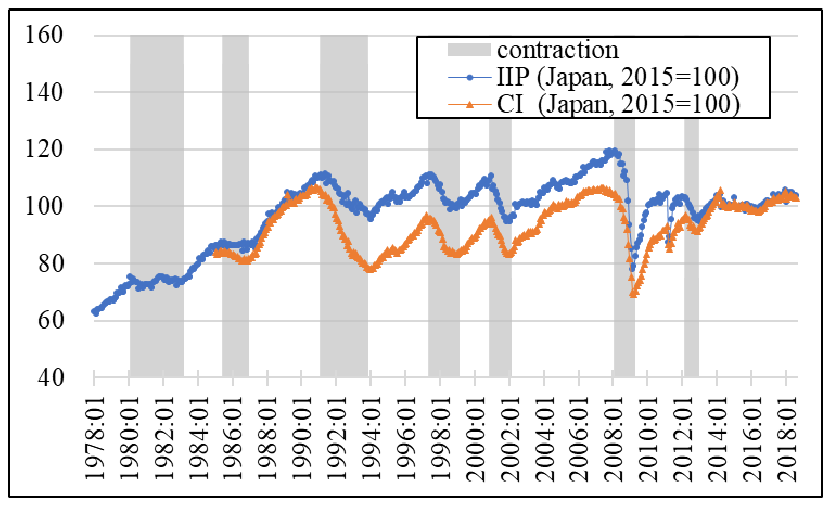}\\
    \end{center}
    \caption{Comparison of Time Series of the Index of Industrial Production and the Composite Index Data for Japan Overall.\newline}
    \begin{spacing}{1.1}
        Note: Timing of peaks and troughs in the index of industrial production data duplicate those of the composite index even though deviations can be large.
    \end{spacing}
    \label{fig:real-filter-IIP-CI}
\end{figure}

It is preferable to employ regional CI data for both countries.
However, we use IIP for Japan because CI data are unavailable for all prefectures.
Figure~\ref{fig:real-filter-IIP-CI} supports our use of IIP data for Japan.
It includes monthly time series of IIP (2015 average $=100$: spanning from January 1978 to August 2018) and CI (2015 average $=100$: spanning from January 1985 to August 2018, which are compiled by the Economic and Social Research Institute (ESRI) of the Cabinet Office) for Japan overall.
The shadowed area corresponds to business contractions, whereas the white area corresponds to business expansions.
Figure~\ref{fig:real-filter-IIP-CI} demonstrates that the timing of peaks and troughs in IIP data duplicate those of CI data even though their deviations from 1991 to 2012 can be large.
In short, IIP data still capture the Japanese BCs adequately.

\section{Measuring Synchronization}
\label{measuring}

We apply three procedures to measure synchronization between two scalar (i.e., one-dimensional) time series.
First, we convert fluctuations in each scalar series into two-dimensional oscillations using the Hilbert transform to identify a ``phase'' at each time. 
Second, we take ``phase differences'' between two cycles as an indicator of their synchronization.
Third, using the phase differences, we calculate a synchronization index proposed by \cite{rosenblum_2001} .

\subsection{Phase Synchronization}
\label{synchro}

\textit{Synchronization} is a phenomenon in which multiple oscillations adjust their individual rhythms through mutual interactions to maintain a constant phase difference for a time.
This phenomenon is strictly called \textit{phase synchronization} and also called \textit{phase-locking} or \textit{frequency entrainment}.
Remark that phase synchronization does not depend on the amplitudes of oscillations.
If their amplitudes are identical, it is called \textit{complete synchronization} (\citet[p.~23]{pikovsky_2001}).
In the present paper, when we use the term synchronization, we mean phase synchronization.

\begin{figure}[p]
    \begin{center}
        \includegraphics{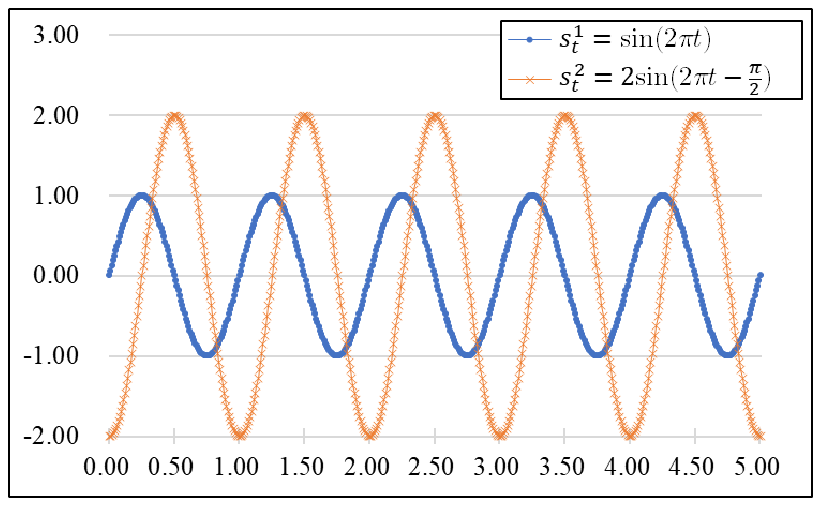}\\
    \end{center}
    \caption{Two Synchronized Time Series with a Temporally Constant Phase Difference.\newline}
    \begin{spacing}{1.1}
        Note: ``$\bullet$'' (blue) represents time series $s^1_t={\sin}(2{\pi}t)$ and ``$\times$'' (orange) represents time series $s^2_t=2{\sin}(2{\pi}t-{\pi}/{2})$. 
        The phase difference is ${\pi}/{2}$ for all $t$.
    \end{spacing}
    \label{fig:time-series-sin-sin2}
\end{figure}

Synchronization is exemplified with the aid of simple oscillators $s^1_t={\sin}(2{\pi}t)$ and $s^2_t=2{\sin}(2{\pi}t - {\pi}/{2})$ in Figure~\ref{fig:time-series-sin-sin2}.
The phases of $s^1_t$ and $s^2_t$ are $2{\pi}t$ and $2{\pi}t-{\pi}/{2}$, respectively,\footnote
{
To be exact, the phase value is restricted to $[-\pi,\pi)$ by taking~$\bmod$ \hspace{-0.21cm} $2\pi$. 
``Phase'' is defined in Section~\ref{Hilbert}. 
}
rendering their phase difference as ${\pi}/{2}$.
Time series $s^1_t$ and $s^2_t$ are synchronized because their phase difference is constant over time.

\subsection{Hilbert Transform and the Instantaneous Phase}
\label{Hilbert}

Phases are crucial in synchronization analysis.
It impossible to extract time-varying amplitudes and phases of BCs just from scalar time series.
However, time-varying amplitudes and phases of BCs make it impossible to extract their information from one-dimensional time series data.
Therefore, we construct a complex-valued time series $\hat{s}_t$ whose real part is actual data $s_t$ and imaginary part $s^H_t$ is generated from $s_t$ via the Hilbert transform: 
\begin{equation}
    \hat{s}_t = s_t + i s^H_t.
    \label{eq:complex}
\end{equation}
The Hilbert transform of $s_t$ is given by
\begin{equation*}
    s_t^H=\frac{1}{\pi}P.V.\int_{-\infty}^{\infty}\frac{s_\tau}{t-\tau}d\tau,
\end{equation*}
where $P.V.$ denotes Cauchy principal value integrals. 
Intuitively, the Hilbert transform provides a phase shift of $-{\pi}/{2}$ radian for every Fourier component of a function. 
For example, the Hilbert transform of $s_t=\cos(2\pi t)$ is $s^H_t=\cos(2\pi t - {\pi}/{2})=\sin(2\pi t)$.

\begin{figure}[p]
    \begin{center}
        \includegraphics[scale=.7]{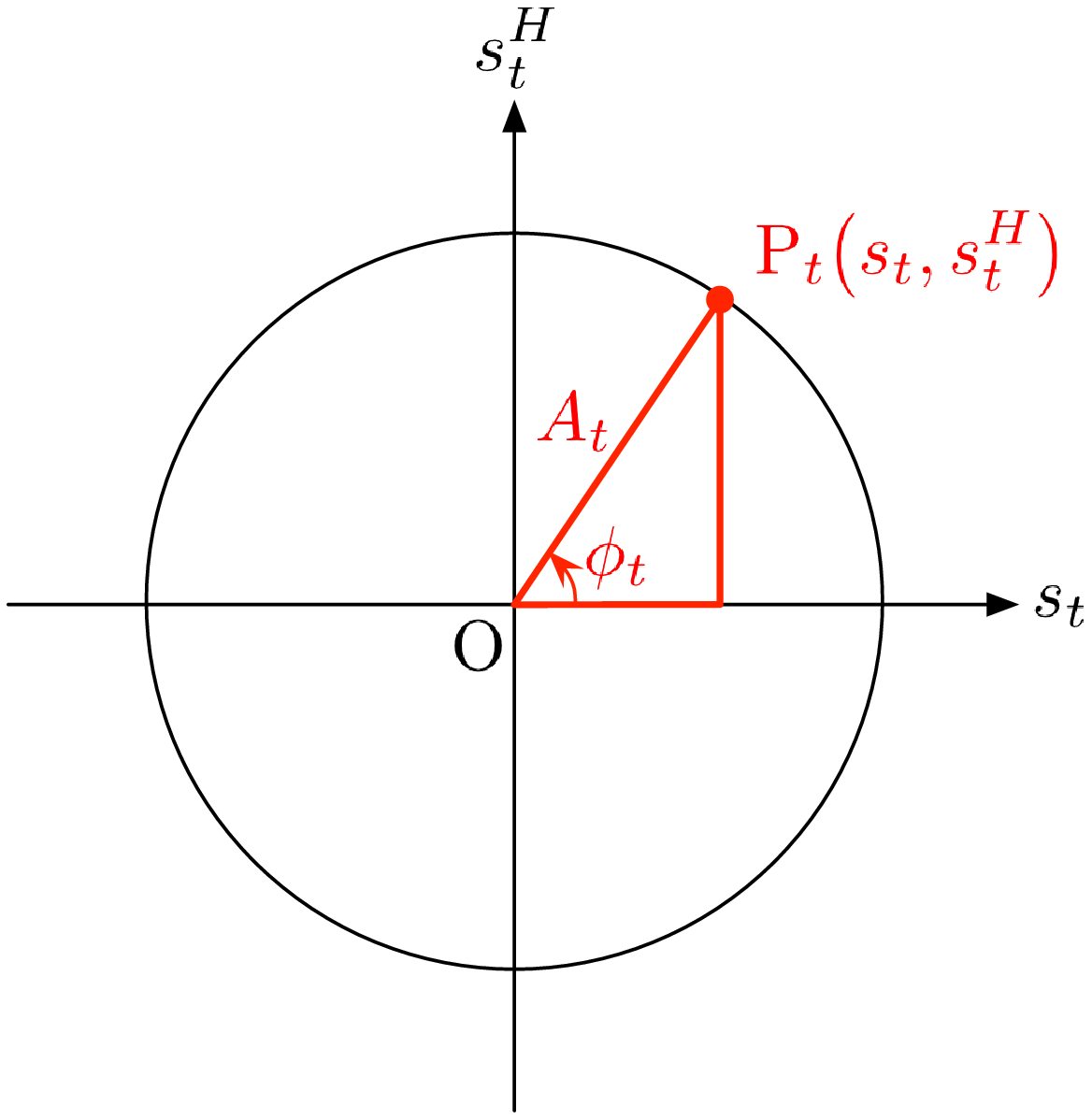}
    \end{center}
    \caption{Constructing a Complex-Valued Time Series $\hat{s}_t$ from One-Dimensional Real Data $s_t$.\newline}
    \begin{spacing}{1.1}
        Note: We construct $\hat{s}_t$ whose real part is actual data $s_t$ and imaginary part $s^H_t$ generated from $s_t$ via the Hilbert transform.
    \end{spacing}
    \label{fig:hilbert}
\end{figure}

Now, we can define a phase (instantaneous) at time $t$ using a point $\text{P}_{t}(s_t,s^H_t)$ on the complex plane as an angle $\phi_t$ that is formed between $\text{OP}_t$ and the horizontal axis in Figure~\ref{fig:hilbert}:
\begin{equation*}
    \phi_t=\left\{\begin{array}{ll}
        \tan^{-1}\left(\displaystyle \frac{s_t^H}{s_t}\right) & (s_t>0),\\
        \tan^{-1}\left(\displaystyle \frac{s_t^H}{s_t}\right)+\pi & (s_t<0).
    \end{array}
    \right.
\end{equation*}
The phase value ranges from $-\pi$ to $\pi$; hence, it can be discontinuous over time.
Using phase $\phi_t$, we can rewrite Equation~\eqref{eq:complex} as
\begin{eqnarray*}
    \hat{s}_t &=& s_t + i s^H_t \nonumber \\
              &=& A_t \cos \phi_t + i A_t \sin \phi_t,
\end{eqnarray*}
where the time-varying amplitude is represented as $A_t = \sqrt{{s_t}^2 + (s^H_t)^2}$.
The earlier discussion assumes that $t$ is continuous; however, we apply the procedure to discrete time series data of CI and IIP in Section \ref{analysis}.

\subsection{Synchronization Index}
\label{index}

To measure degrees of synchronization between two series for the discrete time interval $1 \le i \le W$, we use the synchronization index $\gamma^2 \in [0,1]$ proposed by \cite{rosenblum_2001}:
\begin{equation}
    \gamma^2 = \left( \frac{1}{W} \sum_{i=1}^W \cos\psi_i \right)^2
             + \left( \frac{1}{W} \sum_{i=1}^W \sin\psi_i \right)^2,
    \label{eq:synchronization1}
\end{equation}
where $\psi_i$ denotes the \textit{phase difference} defined by the difference between phases of two time series and $W$ denotes the length of the moving window.
This index $\gamma^2$, also known as the phase-locking value, was first used in the economics literature by \cite{bruzda_2015}.
When $\psi_i$ is nearly constant over time, the value of $\gamma^2$ is close to 1.
This situation is defined as (phase) synchronization.
When $\psi_i$ is chosen randomly from the uniform distribution of $[-\pi,\pi)$, $\gamma^2$ approaches 0 as $W$ increases.

To examine time evolution of $\gamma^2$, we presume $W$ is an odd number of discrete time points, and $\gamma^2_t$ defined below represents the strength of synchronization at time $t$, which corresponds to the temporal center point of the moving window of the length $W$.
Thus, instead of Equation~\eqref{eq:synchronization1}, we calculate for each time $t$
\begin{equation}
    \gamma^2_t = \left( \frac{1}{W} \sum_{i=t-p}^{t+p} \cos\psi_i \right)^2
               + \left( \frac{1}{W} \sum_{i=t-p}^{t+p} \sin\psi_i \right)^2,
    \label{eq:synchronization2}
\end{equation}
where $p={(W-1)}/{2}$ and $0<p<t$.
Throughout the analysis, we set $W=13$ for the U.S. and $W=17$ for Japan because the window's length $W=13$ ($W=17$) equals approximately half of the shortest duration of 28 (36) months of the past BCs of the U.S. (Japan) (see Tables~\ref{table:business-cycle-a} and \ref{table:business-cycle-j}).
A small change in the length of $W$ hardly affects the analysis results (details in Appendix A).
In Equation~\eqref{eq:synchronization2}, the expected values of synchronization index $\gamma^2_t$ with the window's length $W=13$ ($W=17$) is 0.077 (0.059) when $\psi_i$ is chosen randomly from the uniform distribution on $[-\pi,\pi)$.

\section{Analysis}
\label{analysis}

Before applying our method described earlier, we extract recurring patterns from the original time series by employing the band-pass filter based on Fourier series representation (details in Appendix B).
Although other studies have often used BK and CF band-pass filters, we use the  mathematically and computationally simpler Fourier filter.
We reject the BK filter because it uses a moving average and entails excluding a considerable number of data points at both ends to make it perform well.
The time series filtered by Fourier and CF are nearly identical, especially in the timing of peaks and troughs, and qualitative results of synchronization analysis using these two series are identical (details in Appendix C).

When applying the Fourier band-pass filter to the original time series data, 
we must identify the frequency bands corresponding to the time scale
of BCs under consideration.
Tables~\ref{table:business-cycle-a} and~\ref{table:business-cycle-j} list the reference dates for cycles in the U.S. announced by the National Bureau of Economic Research (NBER) and those in Japan by ESRI, from which we identify a band spanning 28--130 months for the U.S. and 36--86 months for Japan, respectively.
We extract time series for the U.S. with the frequency band of that range using the lower cutoff frequency of $k_l=4$ and the upper cutoff frequency of $k_u=18$, which correspond to 126 ($\approx 505/k_l$) and 28 ($\approx 505/k_u$) months, respectively. 
For the notations $k_l$ and $k_u$, see Appendix B.
We also extract time series for Japan with the lower cutoff frequency of $k_l=6$ and the upper cutoff frequency of $k_u=14$, corresponding to 81 ($\approx 488/k_l$) and 35 ($\approx 488/k_u$) months, respectively.
In Appendix D, we perform a robustness check of our analysis against frequency band selection.
Both ends of band-pass-filtered data contain artificial information.
Thus, for the robustness of our results, we eliminate data points at both ends corresponding to the period of the band's highest frequency.

\begin{table}[p]
    \centering
    \caption{Reference Dates for the U.S. Business Cycles Announced by the National Bureau of Economic Research.}
    \vspace{5mm}
    \scalebox{1.0}{
    \begin{tabular}{cccc}
        \hline\noalign{\smallskip}
        Trough & \hspace{1.5mm}Peak & Trough & \hspace{0.5mm}Duration\\
        & & & (months)\\
        \noalign{\smallskip}\hline\noalign{\smallskip}
        1975:03 & 1980:01 & 1980:07 & 64 \\
        1980:07 & 1981:07 & 1982:11 & 28 \\
        1982:11 & 1990:07 & 1991:03 & 100 \\
        1991:03 & 2001:03 & 2001:11 & 128 \\
        2001:11 & 2007:12 & 2009:06 & 91 \\
        2009:06 & 2020:02 & 2020:04 & 130 \\
        \noalign{\smallskip}\hline
    \end{tabular}
    }\\
    \vspace{10mm}
    {Note: Duration of cycles spans 28--130 months.}
    \label{table:business-cycle-a}
\end{table}

\begin{table}[p]
    \centering
    \caption{Reference Dates for Japanese Business Cycles announced by the Economic and Social Research Institute of the Cabinet Office.}
    \vspace{5mm}
    \scalebox{1.0}{
    \begin{tabular}{cccc}
        \hline\noalign{\smallskip}
        Trough & \hspace{1.5mm}Peak & Trough & \hspace{0.5mm}Duration\\
        & & & (months)\\
        \noalign{\smallskip}\hline\noalign{\smallskip}
        1977:10 & 1980:02 & 1983:02 & 64 \\
        1983:02 & 1985:06 & 1986:11 & 45 \\
        1986:11 & 1991:02 & 1993:10 & 83 \\
        1993:10 & 1997:05 & 1999:01 & 63 \\
        1999:01 & 2000:11 & 2002:01 & 36 \\
        2002:01 & 2008:02 & 2009:03 & 86 \\
        2009:03 & 2012:03 & 2012:11 & 44 \\
        2012:11 & 2018:10 & 2020:05 & 90 \\
        \noalign{\smallskip}\hline
    \end{tabular}
    }\\
    \vspace{10mm}
    {Note: Duration of cycles spans 36--86 months.}
    \label{table:business-cycle-j}
\end{table}

\begin{figure}[p]
    \begin{center}
        \subfloat{({\bf u1}) }{\includegraphics[clip, width=0.42\columnwidth,height=0.28\columnwidth]{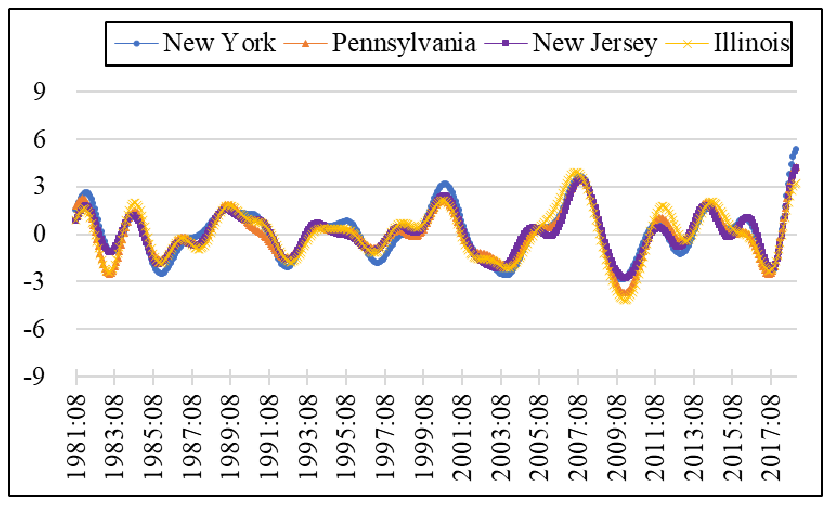}}
        \subfloat{({\bf j1}) }{\includegraphics[clip, width=0.42\columnwidth,height=0.28\columnwidth]{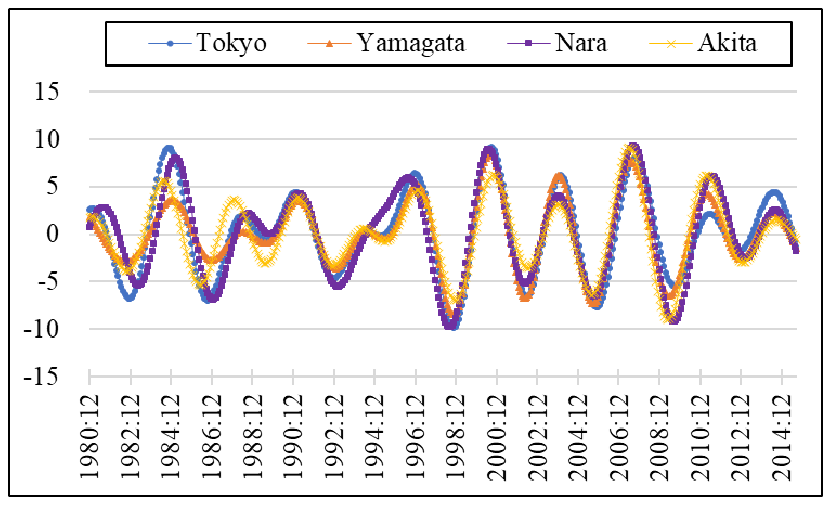}}\\
        \vspace{2mm}
        \subfloat{({\bf u2}) }{\includegraphics[clip, width=0.42\columnwidth,height=0.28\columnwidth]{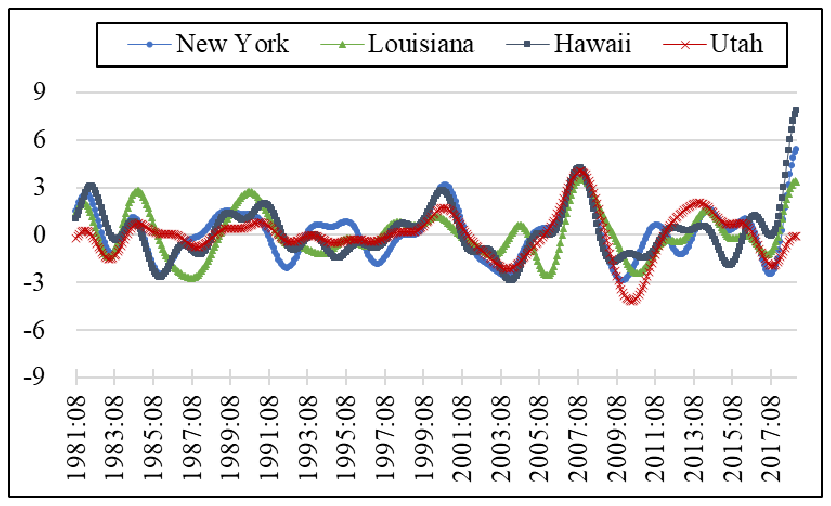}}
        \subfloat{({\bf j2}) }{\includegraphics[clip, width=0.42\columnwidth,height=0.28\columnwidth]{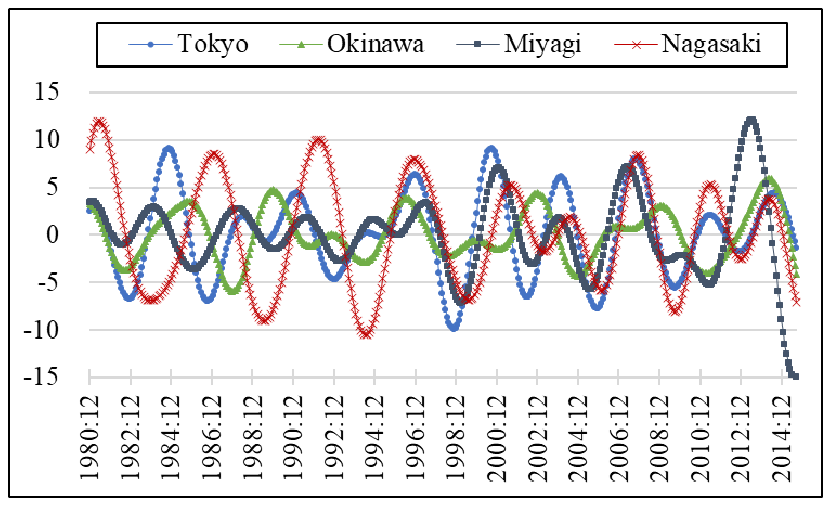}}
    \end{center}
    \caption{Comparisons of Band-Pass-Filtered Time Series of the Composite Index and the Index of Industrial Production Data for Sampled Regions.\newline}
    \begin{spacing}{1.1}
        Note: Comparison of the band-pass-filtered composite index data between New York and other three states whose original data exhibit the greatest synchronization with New York (i.e., Pennsylvania, New Jersey, and Illinois) (u1) and the least synchronization with New York (i.e., Louisiana, Hawaii, and Utah) (u2).
        The upper (lower) cutoff frequency of the band-pass filter is 28 (126) months for (u1) and (u2).
        Comparison of the band-pass-filtered index of industrial production data between Tokyo and other three prefectures whose original data exhibit the greatest synchronization with Tokyo (i.e., Yamagata, Nara, and Akita) (j1) and the least synchronization with Tokyo (i.e., Okinawa, Miyagi, and Nagasaki) (j2).
        The upper (lower) cutoff frequency of the band-pass filter is 35 (81) months for (j1) and (j2).
    \end{spacing}
    \label{fig:band-pass-iip}
\end{figure}

Figure~\ref{fig:band-pass-iip} illustrates the band-pass-filtered time series for sampled regions.
In particular, Figure~\ref{fig:band-pass-iip}~(u1) compares the time series for New York and the same three states as in Figure~\ref{fig:time-series-iip}~(u1).
Meanwhile, Figure~\ref{fig:band-pass-iip}~(j1) compares the time series for Tokyo and the same three prefectures as in Figure~\ref{fig:time-series-iip}~(j1).
The timing of peaks and troughs almost coincides because those regions in (u1) and (j1) are most synchronized.
By contrast, because those regions in (u2) and (j2) are least synchronized with each other, the timing of peaks and troughs is considerably disordered.
These four figures imply periods for which the degree of synchronization between regions in the U.S. and Japan is either high or low.

\begin{figure}[p]
    \begin{center}
        \subfloat{({\bf u1}) }{\includegraphics{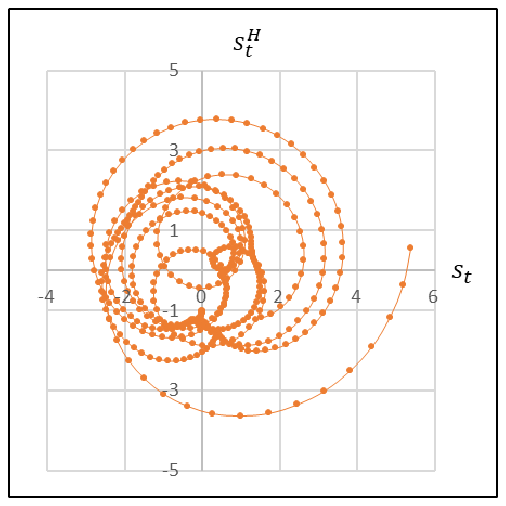}}
        \subfloat{({\bf j1}) }{\includegraphics{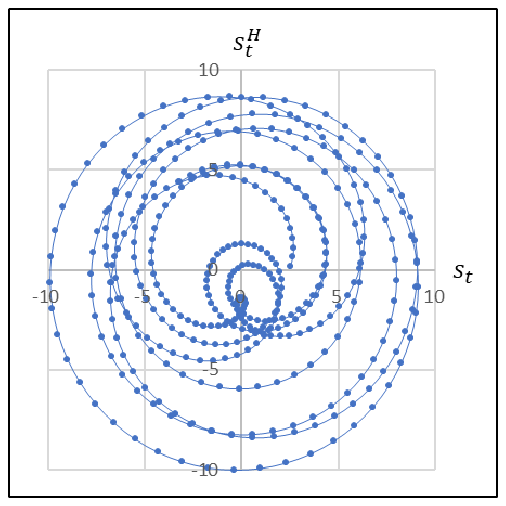}}\\
        \vspace{2mm}
        \subfloat{({\bf u2}) }{\includegraphics{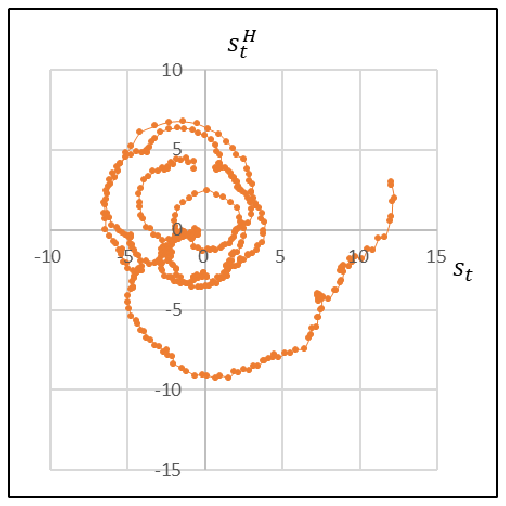}}
        \subfloat{({\bf j2}) }{\includegraphics{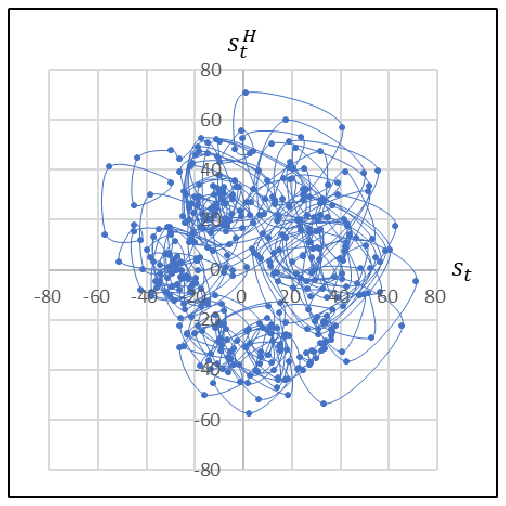}}
    \end{center}
    \caption{Trajectories of the Instantaneous Phase $\text{P}_t(s_t,s^H_t)$ on the Complex Plane.\newline}
    \begin{spacing}{1.1}
        Note: In Figure~\ref{fig:complex}~(u1) (Figure~\ref{fig:complex}~(j1)), the horizontal axis represents the variable $s_t$, that is, the band-pass-filtered composite index (CI) (index of industrial production; IIP) data for New York (Tokyo) with the aforementioned upper and lower cutoff frequencies. 
        In Figure~\ref{fig:complex}~(u2) and (j2), the horizontal axis represents the de-trended but not band-pass-filtered CI (IIP) data for New York (Tokyo).
        In all panels, the vertical axis represents $s^H_t$, that is, the Hilbert-transformed time series of $s_t$.
    \end{spacing}
    \label{fig:complex}
\end{figure}

We next convert fluctuations in each band-pass-filtered scalar time series into two-dimensional oscillations using the Hilbert transform.
Figures~\ref{fig:complex}~(u1) and \ref{fig:complex}~(j1) depict two-dimensional trajectories of the instantaneous phase $\text{P}_t(s_t,s^H_t)$ on the complex plane.
The horizontal axis represents the variable $s_t$, that is, the band-pass-filtered CI (IIP) data for New York (Tokyo) with the aforementioned upper and lower cutoff frequencies, and the vertical axis represents $s^H_t$, that is, the Hilbert-transformed time series of $s_t$.
Trajectories of $\text{P}_t(s_t,s^H_t)$ oscillate around the origin with certain frequencies and amplitudes.
This finding implies that BC fluctuations are adequately extracted using the aforementioned lower and upper cutoff frequencies.

Comparing Figure~\ref{fig:complex}~(u1) with (u2) or Figure~\ref{fig:complex}~(j1) with (j2) uncovers the significance of the band-pass filter and selection of a frequency band.
Figures~\ref{fig:complex}~(u2) and \ref{fig:complex}~(j2) show trajectories of $\text{P}_t(s_t,s^H_t)$ with $s_t$ de-trended but not band-pass-filtered.
Trajectories in Figure~\ref{fig:complex}~(u2) slowly rotate and those in Figure~\ref{fig:complex}~(j2) consist of numerous irregular oscillations.
These imply that time series $s_t$ contains lower (higher)-frequency fluctuations than Figures~\ref{fig:complex}~(u1) and \ref{fig:complex}~(j1) and that BC fluctuations are not adequately extracted.
Moreover, in Figure~\ref{fig:complex}~(j2), trajectories sometimes pass by the origin, suggesting that phase movements exhibit abrupt jumps that may defeat our synchronization analysis.

The converted trajectory on the complex plane via the Hilbert transform allows us to identify the phase of circular oscillations and calculate phase differences between two trajectories as an indicator of the degree of synchronization at each time.
Thus, we can compute the time evolution of synchronization index $\gamma^2_t$ between two trajectories using Equation~\eqref{eq:synchronization2} to gauge the constancy of phase differences.

\begin{figure}[p]
    \begin{center}
        \subfloat{({\bf u1}) }{\includegraphics[clip, width=0.42\columnwidth,height=0.28\columnwidth]{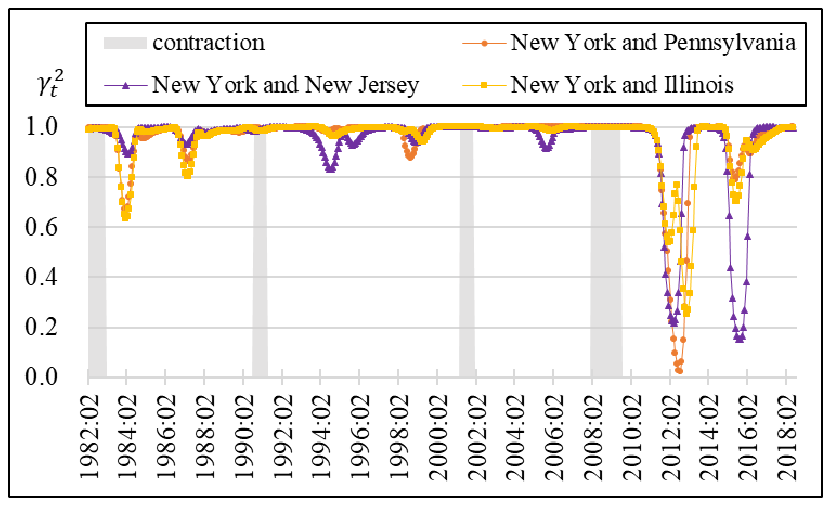}}
        \subfloat{({\bf j1}) }{\includegraphics[clip, width=0.42\columnwidth,height=0.28\columnwidth]{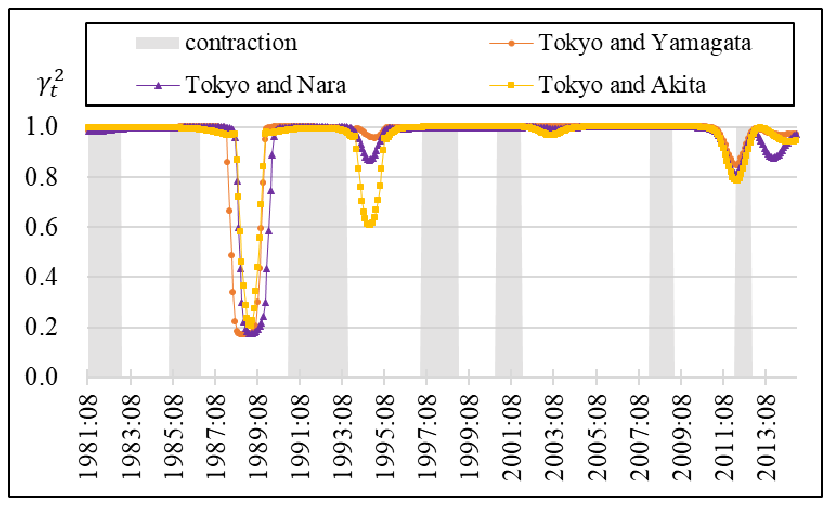}}\\
        \vspace{2mm}
        \subfloat{({\bf u2}) }{\includegraphics[clip, width=0.42\columnwidth,height=0.28\columnwidth]{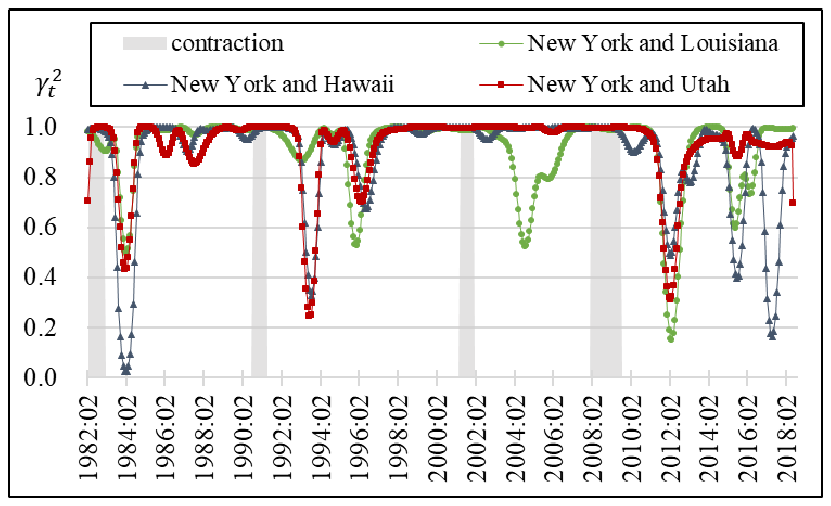}}
        \subfloat{({\bf j2}) }{\includegraphics[clip, width=0.42\columnwidth,height=0.28\columnwidth]{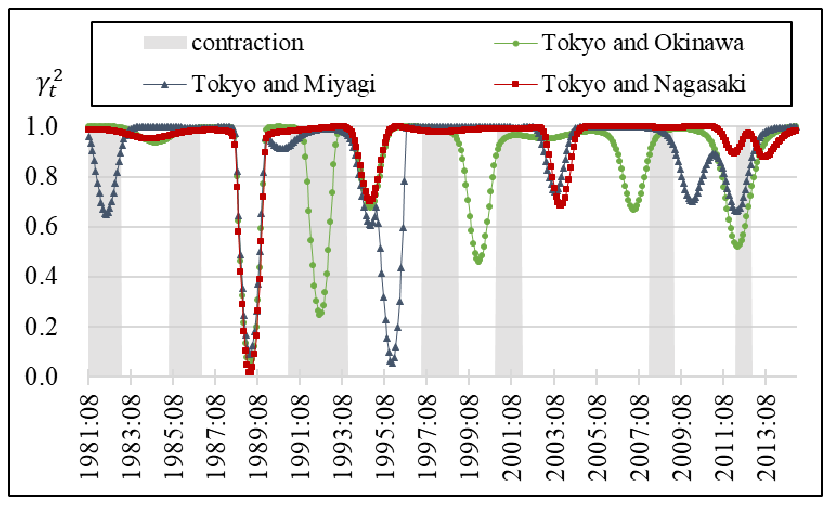}}
    \end{center}
    \caption{Time Evolution of the Synchronization Index $\gamma^2_t$ between Sampled Regions.\newline}
    \begin{spacing}{1.1}
        Note: Time Evolution of $\gamma^2_t$ between New York and other three states whose original composite index data exhibit the greatest synchronization with New York (i.e., Pennsylvania, New Jersey, and Illinois) (u1) and the least synchronization with New York (i.e, Louisiana, Hawaii, and  Utah) (u2). 
        Time Evolution of $\gamma^2_t$ between Tokyo and other three prefectures whose original index of industrial production data exhibit the greatest synchronization with Tokyo (i.e., Yamagata, Nara, and Akita) (j1) and the least synchronization with Tokyo (i.e., Okinawa, Miyagi, and Nagasaki) (j2).
    \end{spacing}
    \label{fig:time-sync-ind-jp-pref}
\end{figure}

Figure~\ref{fig:time-sync-ind-jp-pref} illustrates time evolution of synchronization index $\gamma^2_t$ between sampled regions.
States in Figures~\ref{fig:time-sync-ind-jp-pref}~(u1) and \ref{fig:time-sync-ind-jp-pref}~(u2) correspond to those in Figures~\ref{fig:time-series-iip}~(u1) and \ref{fig:time-series-iip}~(u2), respectively.
Likewise, prefectures in Figures~\ref{fig:time-sync-ind-jp-pref}~(j1) and \ref{fig:time-sync-ind-jp-pref}~(j2) correspond to those in Figures~\ref{fig:time-series-iip}~(j1) and \ref{fig:time-series-iip}~(j2), respectively.
Although states and prefectures in Figures~\ref{fig:time-sync-ind-jp-pref}~(u1) and (j1) belong to the most synchronized group with New York and Tokyo, the figures display some intervals during which $\gamma^2_t$ takes low values.
Figures~\ref{fig:time-sync-ind-jp-pref}~(u2) and \ref{fig:time-sync-ind-jp-pref}~(j2) depict states and prefectures that are least synchronized with New York and Tokyo, so that the degree of synchronization is low compared with that in Figures~\ref{fig:time-sync-ind-jp-pref}~(u1) and \ref{fig:time-sync-ind-jp-pref}~(j1).
These figures imply that synchronization is generally high during most periods and tends to decline concurrently almost during BC expansions (white areas).

To scrutinize the degree of BC synchronization between regions, we calculate 
1,225 (=$_{50}\mathrm{C}_2$) series of $\gamma^2_t$ for all two-tuples between the 50 states in the U.S., and 1,081 (=$_{47}\mathrm{C}_2$) series of $\gamma^2_t$ for all two-tuples between Japanese 47 prefectures.
By $R(\gamma^2_t \ge r)$, we denote the ratio of two-tuples for which $\gamma^2_t$ exceeds or equals the threshold $r$ at each time $t$.
By definition, $R(\gamma^2_t \ge r) \in [0,1]$.
The larger the portion of prefectures synchronized, the greater the value of $R(\gamma^2_t \ge r)$.

\begin{figure}[p]
    \begin{center}
        \subfloat{({\bf u})}{\includegraphics{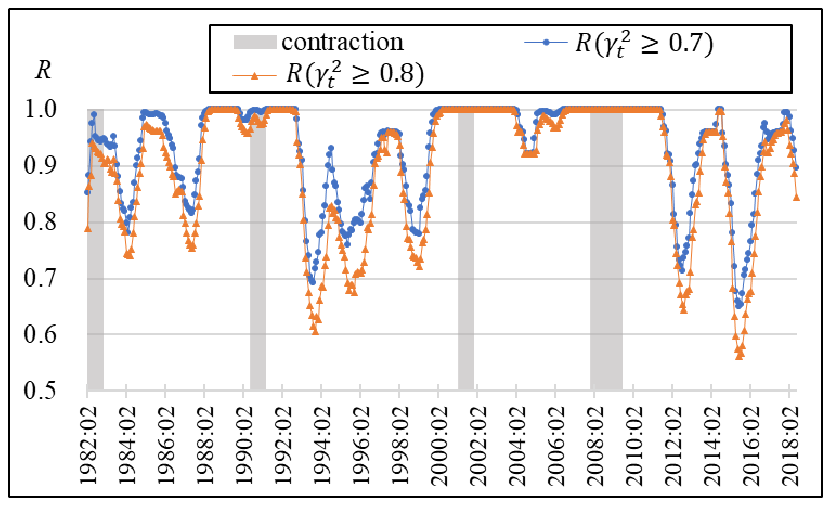}}\\
        \vspace{2mm}
        \subfloat{({\bf j}) }{\includegraphics{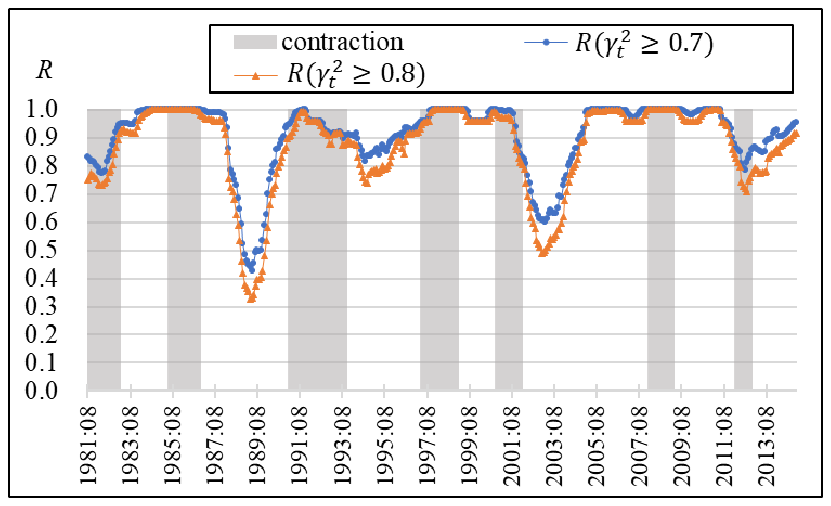}}
    \end{center}
    \caption{Time Evolution of the Ratio $R(\gamma^2_t \ge r)$ for $r=0.7$ and $0.8$.\newline}
    \begin{spacing}{1.1}    
        Note: By $R(\gamma^2_t \ge r)$, we denote the ratio of 2-tuples for which $\gamma^2_t$ takes a value greater than or equal to the threshold $r$ at each time $t$.
    \end{spacing}
    \label{fig:combinations}
\end{figure}

Figure~\ref{fig:combinations} illustrates the time evolution of $R(\gamma^2 \ge r)$ for $r = 0.7$ and $0.8$.
It implies that, in both countries, $R(\gamma^2_t \ge r)$ is inclined to be low during expansions (white areas), whereas it is inclined to be high during contractions (shadowed areas).
These observations support the hypothesis that the degree of synchronization between regional BCs increases during contractions and decreases during expansions.

\begin{figure}[p]
    \begin{center}
        \includegraphics[width=\columnwidth]{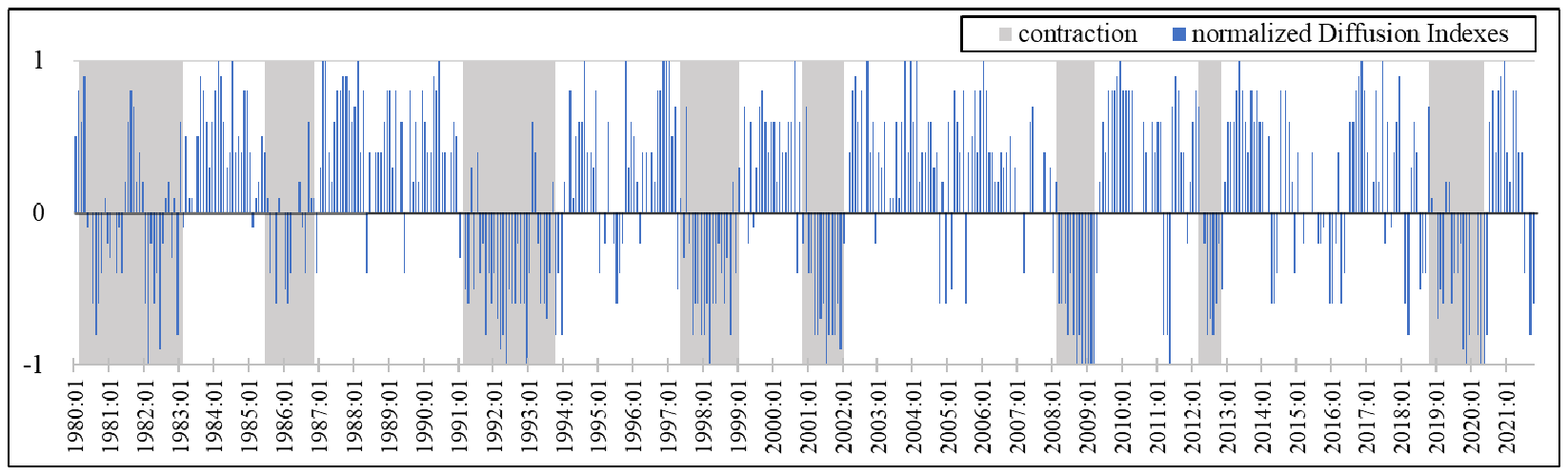}
    \end{center}
    \caption{Time Evolution of the Normalized Diffusion Index in Japan.\newline} 
    \begin{spacing}{1.1}    
        Note: The original diffusion index (DI) data $x_t$ is normalized as $(x_t-50)/50$.
        The normalized DI tends to be positive in expansions and negative in contractions.
    \end{spacing}
    \label{fig:NDI}
\end{figure}

However, Figure~\ref{fig:combinations}~(j) reveals two discrepancies between our results and the hypothesis concerning Japan.
One is that $R(\gamma^2_t \ge r)$ shows relatively low values for the contraction from 2012:03 to 2012:11.
This is because a band-pass filter may fail to extract an adequate trajectory if the duration of an expansion or contraction is too short compared to the period corresponding to the cutoff frequency.
The contraction in question is actually a short period of 8 months.
The other is that $R(\gamma^2_t \ge r)$ shows relatively low value during the contraction from 1981:08 to 1983:02.
This is because observations may disparage our hypothesis during periods when expansions and contractions coexist, that is, when Japan's economy does not expand or contract unidirectionally.
To see this in detail, we inquire into the Diffusion Index (DI) of coincident indicators.
Figure~\ref{fig:NDI} illustrates the time evolution of the normalized DI data in Japan.
The original DI data $x_t$ that takes values from $0$ to $100$ is normalized as $(x_t-50)/50$; hence, the normalized DI tends to be positive in expansions and negative in contractions.
During the contraction from 1980:02 to 1983:02, the normalized DI moves back and forth between positive and negative regions several times, implying some expansions during a BC contraction.
Therefore, in Figure~\ref{fig:combinations}, $R(\gamma^2_t \ge r)$ during that contraction period exhibits relatively low values.
In a nutshell, our observations might deviate from our hypothesis because our method captures interims of expansions within contractions and vice versa sensitively.
This does not constitute a defect in our method.

Finally, we offer a conjecture about why our hypothesis holds, that is, why the degree of synchronization between regional BCs increases during contractions and decreases during expansions.
Under prospect theory (\citealp{kahneman_1979}), people prefer avoiding losses over acquiring equivalent gains.
This suggests industrial firms behave asymmetrically when BCs enter contractions or expansions.
When entering a contraction, firms trim production to avoid losses, and that behavior synchronizes well.
When entering an expansion, some firms step up production and others do not, and their behavior tends not to synchronize well.
Thus, loss-averse behavior by firms engenders synchronization in production during contractions.

\section{Conclusion}
\label{conclusion}

We investigate CI data for all 50 states in the U.S. and IIP data for all 47 prefectures in Japan from the viewpoint of regional BC synchronization. 
Using a method distinguished in nonlinear sciences to analyze synchronization between data series, we converted one-dimensional time series into two-dimensional circular oscillations via the Hilbert transform.
Our quantitative results indicate an increase (decrease) in synchronization of regional BCs during contractions (expansions) throughout the period under study.
Such asymmetry between the contraction and expansion phases of a BC will contribute our better understanding of the phenomenon of BCs.
Among other things, our results provide important information to policymakers. 
This is because, during contractions, regional BCs tend to be coherent, so counter-cyclical fiscal and monetary policies need to be implemented as quickly as possible. 
In contrast, during expansions, regional cycles are less coherent and therefore counter-cyclical policies are less urgent.
Furthermore, as Figure~\ref{fig:combinations}(u) shows, the degree of synchronization can rise and fall significantly several times during a single expansionary period. 
This implies that the economy is not monotonically expanding in the period assigned as an expansionary period by the business cycle reference date. 
If this is the case, our method may allow us to subdivide the expansionary and recessionary periods.

One limitation of our results is that our method concentrates on a specific frequency band and may fail to extract a good trajectory if the duration of an expansion or contraction is too short compared to the period corresponding to the cutoff frequency of the band-pass filter.

Future research should generalize our findings by applying our method to regional BCs in other countries and even to cross-border regions.
In particular, synchronization of BCs in EU countries, which were excluded from the analysis in this study, is of primary importance.
Furthermore, it would be interesting to analyze how the impact of COVID-19 has brought about changes in the appearance of the regional BCs compared to prior years. Incidentally, \cite{dehaan_2022} found that the impact of COVID-19 was strengthened the synchronization of BCs in EU countries, but with large differences in amplitude.
It would also be useful to reexamine our hypothesis via different methods such as wavelet analysis, and construct a macroeconomic dynamical model with loss-averse behavior of firms to explain our hypothesis.

\newpage
\section*{Acknowledgements}
The authors would like to thank the anonymous referees for helpful comments.
This work was partly supported by JST PRESTO (JPMJPR16E5), JSPS KAKENHI (17K05360, 19K01593, 19KK0067, and 21K18584), Tokio Marine Kagami Memorial Foundation.

\newpage
\section*{Appendix A: Robustness against the Length of the Moving Window}

When calculating the synchronization index $\gamma^2$, we employ the moving window $W$ in Equation~\eqref{eq:synchronization2} with a length of $13$ for the U.S. CI data and that with $17$ for Japan's IIP data. 
We chose these lengths of $W$ to be approximately half of the shortest duration of the past BCs.
Here we discuss the robustness of our analysis against the selection of the window's length.

\begin{figure}[p]
    \begin{center}
        \subfloat{({\bf u}) }{\includegraphics[clip, width=0.42\columnwidth,height=0.28\columnwidth]{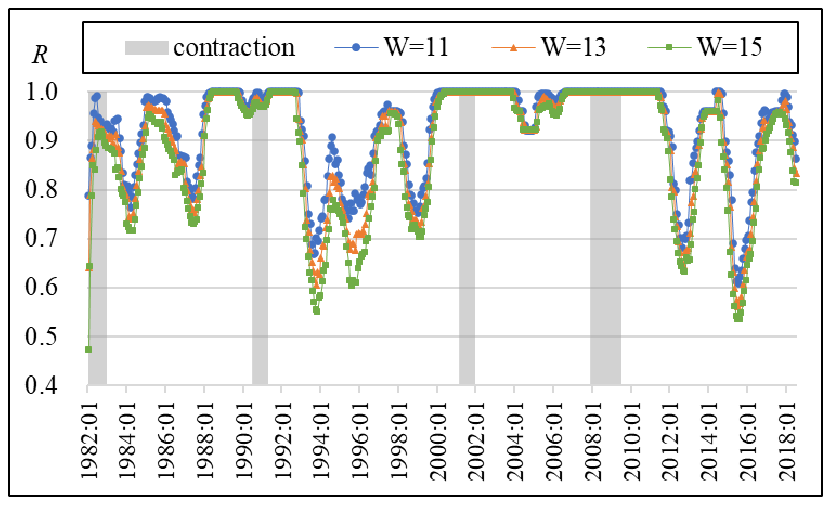}}
        \subfloat{({\bf j}) }{\includegraphics[clip, width=0.42\columnwidth,height=0.28\columnwidth]{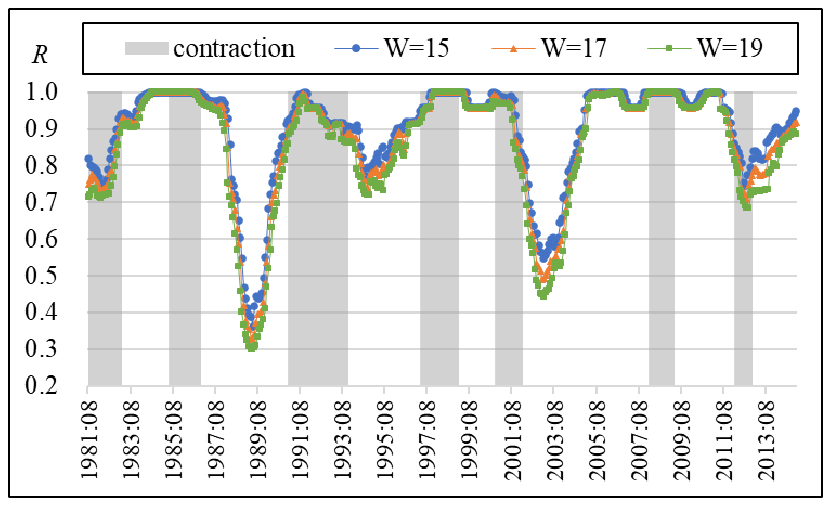}}
    \end{center}
    \caption{Time Evolution of the Ratio $R(\gamma^2_t \ge r)$ for Different Window's Length with Respect to $r=0.8$.\newline}
    \begin{spacing}{1.1}
        Note: Time series of $R(\gamma^2_t \ge 0.8)$ for the U.S. composite index data with respect to window's length $W=11$, $W=13$, and $W=15$ (u).
        Time series of $R(\gamma^2_t \ge 0.8)$ for Japan's index of industrial production data with respect to $W=15$, $W=17$, and $W=19$ (j).
    \end{spacing}
    \label{fig:combinations-win}
\end{figure}

Figure~\ref{fig:combinations-win} compares time evolution of the ratio $R(\gamma^2_t \ge 0.8)$ regarding longer and shorter windows than $W=13$ for the U.S. (u) and $W=17$ for Japan (j).
These panels indicate almost no difference in the analysis results for both countries even if the length of the window varies.

\newpage
\section*{Appendix B: Fourier Filter}

We briefly review the Fourier series of a function $f$.  
For simplicity, let $f$ be a real-valued continuous periodic function on $[0,L)$.
The function $f$ can be represented as a Fourier series:
  \begin{equation}
  f(x)=\frac{a_0}{2}+\displaystyle\sum^\infty_{k=1} \left(a_k \cos \left(\frac{2\pi kx}{L}\right) + b_k \sin \left(\frac{2\pi kx}{L}\right)\right),
  \label{eq:fourier}
  \end{equation}
  where
  \begin{eqnarray*}
    a_k &=& \frac{1}{L}\int^{L}_{0} f(x) \cos\left(\frac{2\pi k x}{L}\right)~dx~(k=0,1,2,3,\ldots),\\
    b_k &=& \frac{1}{L}\int^{L}_{0} f(x) \sin\left(\frac{2\pi k x}{L}\right)~dx~(k=1,2,3,\ldots).
  \end{eqnarray*}
We can obtain a Fourier series for a more general class of function $f$ \citep[see, e.g., ][]{korner_1989}.

By taking a partial sum in Equation~\eqref{eq:fourier} we can create a band-pass-filtered periodic function $\tilde{f}$ 
using a band $[k_l,k_u]$ with the lower and upper cutoff frequencies of $k_l$ and $k_u$ $(0\le k_l \le k \le k_u)$ from a given function $f$: 
  \begin{equation*}
    \tilde{f}(x)=\displaystyle\sum^{k_u}_{k=k_l} \left(a_k \cos \left(\frac{2\pi kx}{L}\right) + b_k \sin \left(\frac{2\pi kx}{L}\right)\right).
  \end{equation*}

\newpage
\section*{Appendix C: Robustness against Filter Selection}

\begin{figure}[p]
    \begin{center}
        \subfloat{({\bf u})}{\includegraphics{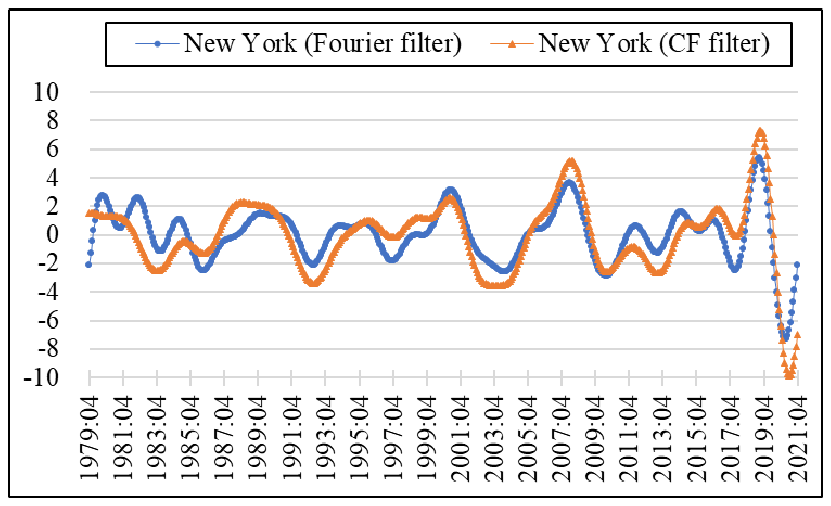}}\\
        \vspace{2mm}
        \subfloat{({\bf j}) }{\includegraphics{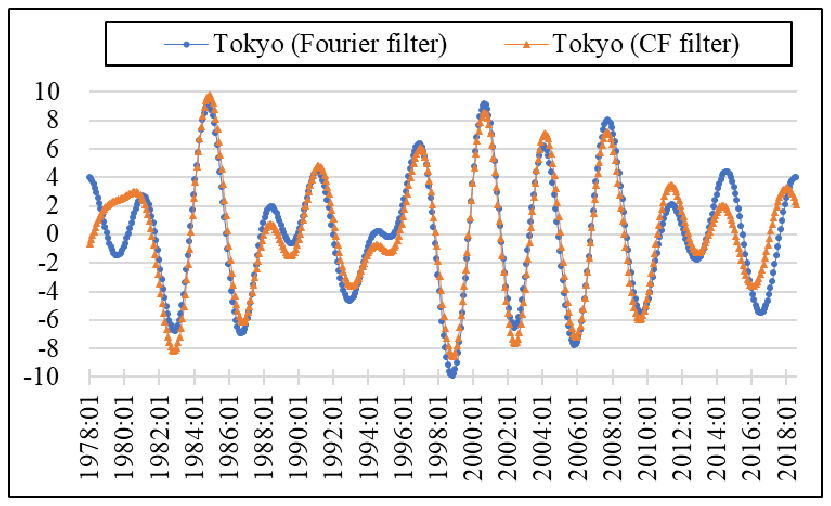}}
    \end{center}
    \caption{Comparison of Band-Pass-Filtered Time Series of the Composite Index and the Index of Industrial Production by Fourier and Christiano--Fitzgerald Filters.\newline}
    \begin{spacing}{1.1}    
        Note: The band-pass filters' upper (lower) cutoff frequency is 28 (126) months in the upper panel. 
        The band-pass filters' upper (lower) cutoff frequency is 35 (81) months in the lower panel.
    \end{spacing}
    \label{fig:fourier-cf}
\end{figure}

We see that qualitative results of our analysis using two different band-pass filters, Fourier and CF, are identical.
Figure~\ref{fig:fourier-cf} shows the filtered time series of CI in New York and those of IIP in Tokyo by both Fourier and CF filters.
The band-pass filters' upper (lower) cutoff frequency of the band-pass filters is 28 (126) months in the upper panel. 
The band-pass filters' upper (lower) cutoff frequency of the band-pass filters is 35 (81) months in the lower panel.
Therefore, the two band-pass filtered time series are similar except for both ends of the data period.
Both ends of band-pass-filtered data contain artificial information; thus, we eliminate data points at both ends corresponding to the period of the highest frequency of the band for the robustness of results.

\begin{figure}[p]
    \begin{center}
        \subfloat{({\bf u1}) }{\includegraphics[clip, width=0.42\columnwidth,height=0.28\columnwidth]{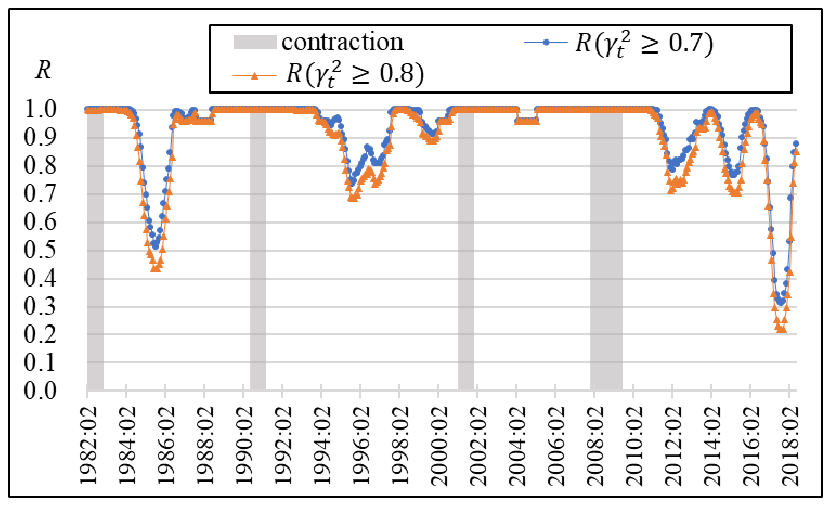}}
        \subfloat{({\bf j1}) }{\includegraphics[clip, width=0.42\columnwidth,height=0.28\columnwidth]{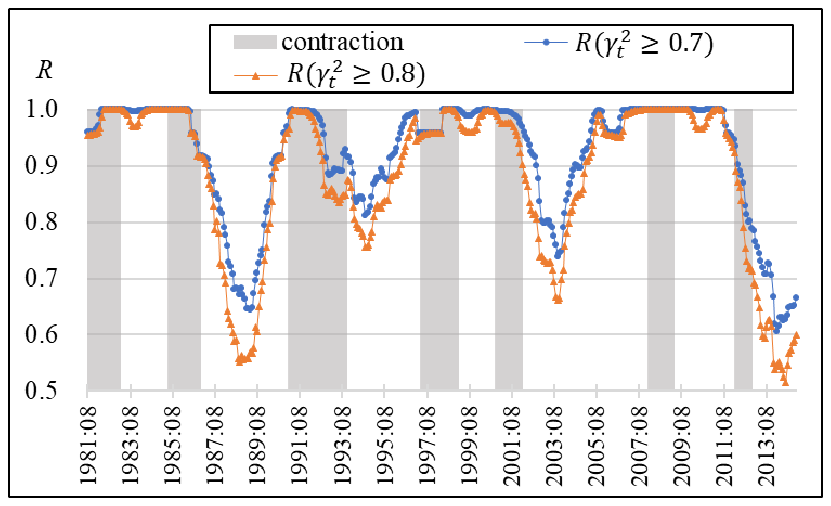}}\\
        \vspace{2mm}
        \subfloat{({\bf u2}) }{\includegraphics[clip, width=0.42\columnwidth,height=0.28\columnwidth]{combinations-1225-a.eps}}
        \subfloat{({\bf j2}) }{\includegraphics[clip, width=0.42\columnwidth,height=0.28\columnwidth]{combinations-1081-j.eps}}\\
    \end{center}
    \caption{Time Evolution of the Ratio $R(\gamma^2_t \ge r)$ for $r=0.7$ and $0.8$ of Band-Pass Filtered Time Series by CF (upper) and Fourier series (lower).\newline}
    \begin{spacing}{1.1}
        Note: The lower panels are reprints from Figure~\ref{fig:combinations}.
    \end{spacing}
    \label{fig:combinations-CF}
\end{figure}

Figure~\ref{fig:combinations-CF} depicts the time evolution of the ratio $R(\gamma^2_t \ge r)$ for $r=0.7$ and $0.8$ of band-pass-filtered time series by CF filter (upper) and those by Fourier filter (lower).
The lower panels are reprints from Figure~\ref{fig:combinations}.
The qualitative results of the analysis are almost the same for both filters, although some differences exist in detail.
Therefore, the filter selection robustness follows.

\newpage
\section*{Appendix D: Robustness against Frequency Band Selection}

When applying the Fourier band-pass filter, we employ the frequency band spanning 28--126  ($k_l=4, k_u=18$) months for the U.S. CI data and 35--81 months ($k_l=6, k_u=14$) for Japan's IIP data.
Here we discuss the robustness of our analysis against frequency band selection.

\begin{figure}[p]
    \begin{center}
        \subfloat{({\bf u1}) }{\includegraphics[clip, width=0.42\columnwidth,height=0.28\columnwidth]{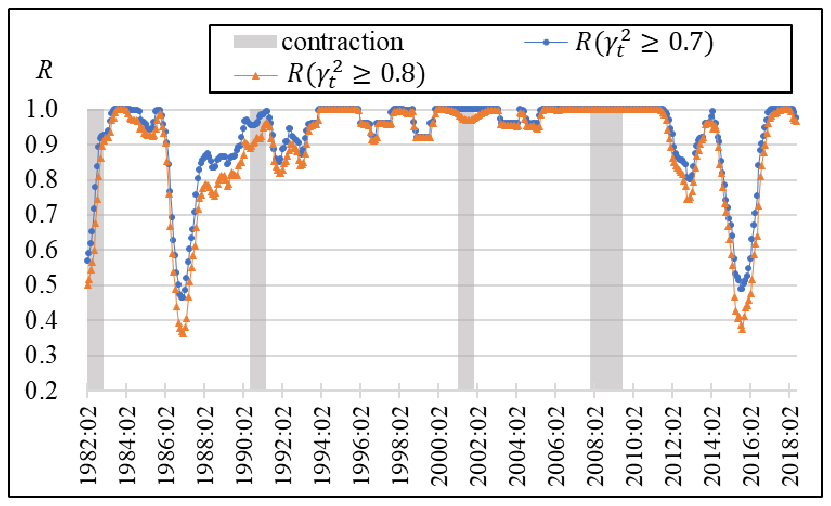}}
        \subfloat{({\bf j1}) }{\includegraphics[clip, width=0.42\columnwidth,height=0.28\columnwidth]{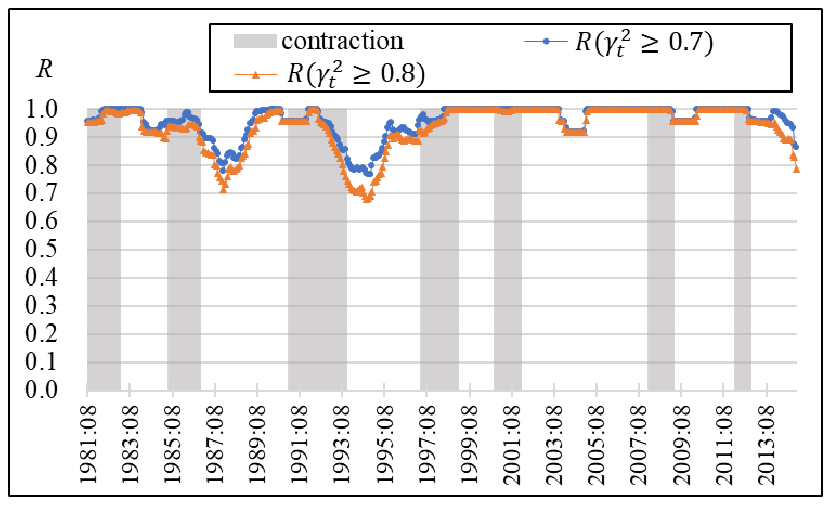}}\\
        \vspace{2mm}
        \subfloat{({\bf u2}) }{\includegraphics[clip, width=0.42\columnwidth,height=0.28\columnwidth]{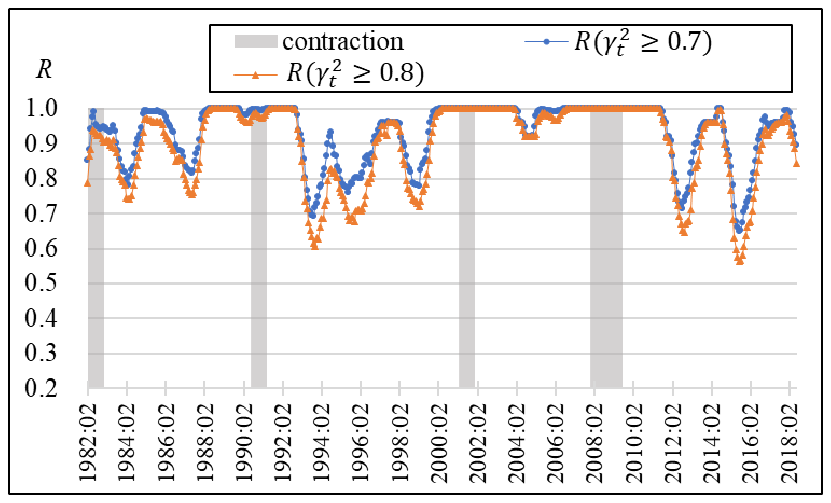}}
        \subfloat{({\bf j2}) }{\includegraphics[clip, width=0.42\columnwidth,height=0.28\columnwidth]{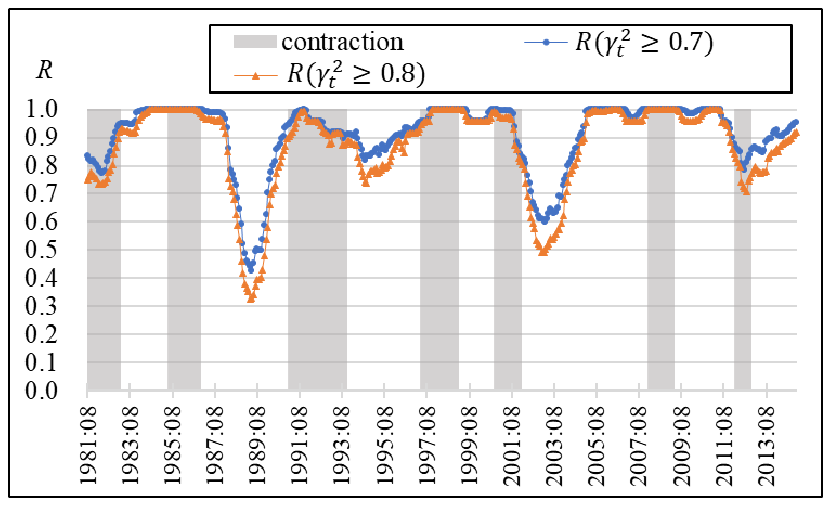}}\\
        \vspace{2mm}
        \subfloat{({\bf u3}) }{\includegraphics[clip, width=0.42\columnwidth,height=0.28\columnwidth]{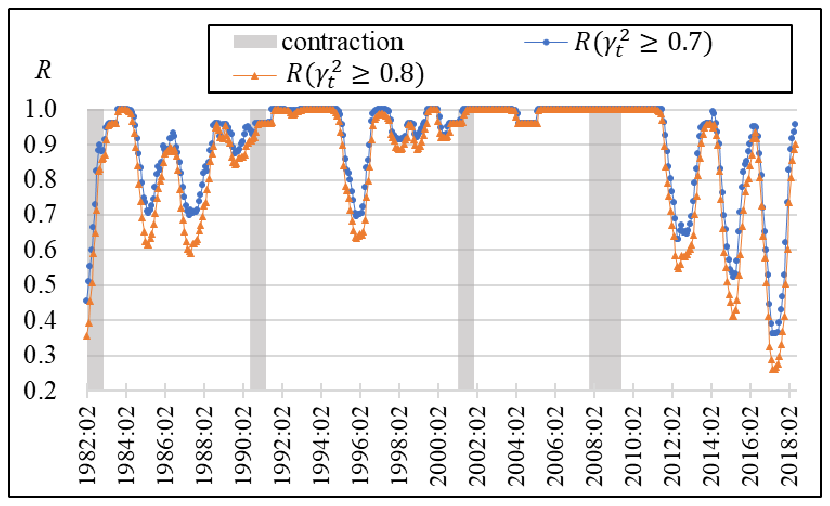}}
        \subfloat{({\bf j3}) }{\includegraphics[clip, width=0.42\columnwidth,height=0.28\columnwidth]{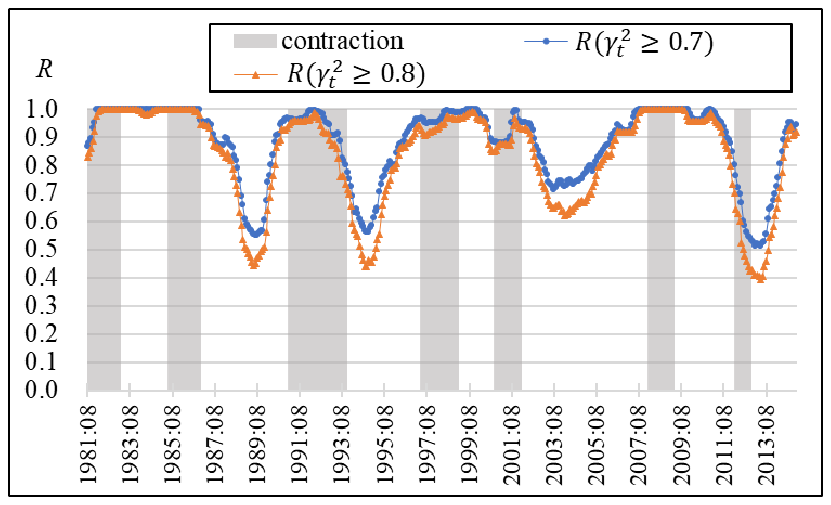}}
    \end{center}
    \caption{Time Evolution of the Ratio $R(\gamma^2_t \ge r)$ for Different Frequency Bands with respect to $r=0.7$ and $0.8$.\newline}
    \begin{spacing}{1.1}
        Note: Time series of $R(\gamma^2_t \ge r)$ for the U.S. composite index data with respect to frequency bands spanning 30 to 101 months ($k_l=5, k_u=17$) (u1), 28 to 126 months ($k_l=4, k_u=18$) (u2), and 27 to 168 months ($k_l=3, k_u=19$) (u3).
        Time series of $R(\gamma^2_t \ge r)$ for Japan's index of industrial production data with respect to frequency bands spanning 38 to 70 months ($k_l=7, k_u=13$) (j1), 35 to 81 months ($k_l=6, k_u=14$) (j2), and 33 to 98 months ($k_l=5, k_u=15$) (j3).
        The two panels in the middle are reprints from Figure~\ref{fig:combinations}.
    \end{spacing}
    \label{fig:combinations-bands}
\end{figure}

Figure~\ref{fig:combinations-bands} illustrates time evolution of the ratio $R(\gamma^2_t \ge r)$ with respect to $r=0.7$ and $0.8$ for different frequency bands.
The two panels in the middle are reprints from Figure~\ref{fig:combinations}, which depict time series of $R(\gamma^2_t \ge r)$ with respect to frequency bands spanning 28 to 126 months ($k_l=4, k_u=18$) (u2) and 35 to 81 months ($k_l=6, k_u=14$) (j2).
The two upper panels correspond to a shorter frequency band spanning 30 to 101 months ($k_l=5, k_u=17$) (u1) and 38 to 70 months ($k_l=7, k_u=13$) (j1), and the two lower panels correspond to a longer frequency band spanning 27 to 168 months ($k_l=3, k_u=19$) (u3) and 33 to 98 months ($k_l=5, k_u=15$) (j3).

Comparing these panels vertically, we show almost identical qualitative results for both U.S. and Japan, although the shape of the graphs varies to some extent depending on the choice of the frequency band.

\end{document}